# Dispersion of Resonant Raman Peaks of CO and OH in $SnO_2$, $Mo_{1-x}Fe_xO_2$ Thin Films and $SiO_2$ bulk glass

B. N. Raja Sekhar[1], R. J. Choudhary[2], D.M. Phase[2] and Shailendra Kumar[*3]

[1]Spectroscopy Division, Bhabha Atomic Research Centre, Mumbai-40085, India

[2]UGC-DAE Consortium for Scientific Research, Indore-452017, India

[3]Laser Materials Development & Devices Division,

Raja Ramanna Centre for Advanced Technology, Indore-452013, India

……………………………………………………………………………

E-mail address: Shailendra Kumar - shail@cat.ernet.in

Phone: +91-731-2488377

Fax: +91-731-2488300

**Abstract**

Resonance Raman (RR) peaks of *CO* and *OH* stretching modes and their higher harmonics have been observed superimposed on photoluminescence (PL) spectrum of $SnO_2$ thin films. Commercial fluorine doped $SnO_2$ thin films deposited by sputtering on glass and $SnO_2$ thin films deposited on Si by laser ablation have been studied. The dispersions of CO and OH stretching RR modes are ~ 600 $cm^{-1}$/eV and 800 $cm^{-1}$/eV respectively. The dispersion of the third harmonic of CO stretching mode is ~ 2000 $cm^{-1}$/eV. Similar dispersion of RR peak of *CO* stretching modes and higher harmonics superimposed on PL spectra has been observed in $Mo_{1-x}Fe_xO_2$ thin films and $SiO_2$ bulk glass. Large dispersion of RR peaks seems to be a common property of oxides with impurities of *CO* and *OH*.

**Introduction**:

Photoluminescence (PL) studies are routinely used to characterize photonic materials. Interest in research activities in oxide materials in bulk, thin film and nano-structure geometries has drawn attention of many research groups for their uses in UV and VUV ranges. We have observed sharp peaks superimposed on PL spectra of $SnO_2$, $Mo_{1-x}Fe_xO_2$ thin films, bulk fused and ordinary $SiO_2$ glass. These superimposed peaks are due to resonance Raman (RR) of vibrational modes of CO and OH impurities present in these oxides. Common results related with fundamental and higher harmonics of RR peaks and their dispersion with respect to incident energy are presented in this letter.

**Experiment**

Commercial F doped $SnO_2$ thin films deposited by sputtering on glass and $SnO_2$ thin films deposited on Si by laser ablation have been used for PL study. The ceramic target used for SnO2 film growth by pulsed laser ablation was prepared by standard solid state reaction technique. A KrF excimer laser source ($\lambda$ = 248nm, pulse width = 20 ns) was used to ablate the target. Substrate was cleaned by chemical method thoroughly and the chamber was evacuated at a base pressure of $2 \times 10^{-6}$ Torr before deposition. Depositions were performed at a substrate temperature of $700^0$ C and oxygen partial pressure of $1 \times 10^{-4}$ Torr, while target to substrate distance was 5 cm. The laser energy density and pulse repetition rate were kept at 1.8J/cm2 and 10 Hz, respectively. The samples were cooled in the same pressure as used during the deposition, at the rate of 20 °C/min. Thicknesses of films were 200 nm as determined using Talystep profilometer. These thin films are polycrystalline. $Mo_{1-x}Fe_xO_2$ thin films were also deposited by laser ablation on sapphire substrates in a similar process. PL studies were carried out using unpolarized excitation source as the incident ray. Outgoing scattered rays were recorded using back reflection geometry and spectra were recorded with incremental step of 0.1 nm. The photoluminescence and resonance Raman spectral data were obtained using a spectrofluorometer (FluoroMax –3, Jobin Vyon). The background excitation source used for this purpose was a 150W xenon, continuous output, ozone-free lamp. A calibrated photodiode and a R928P photomultiplier are used as excitation and emission detectors. All PL measurements have been done at room temperature in air. Excitation energies have been varied from 2.5eV to 5eV. Resonance Raman measurements have been done on a photoluminescence experimental setup and scattered light is not measured in the range between (incident wavelength) and (incident wavelength plus 15 nm) due to large Rayleigh scattering. Scattered light is measured for shift energy more than 1800 cm$^{-1}$. Hence, Raman scattering results for shift energy less than 1800cm$^{-1}$ are not given in this manuscript. Normal Raman spectrum of $SnO_2$ thin films, measured on a Raman experimental setup, is similar to published results in the literature [1].

**Results and Discussion**

Emiroglu et. al. studied vibrational modes of $OH, CO, H_3O^+$ and $H_5O_2^+$ attached to $SnO_2$ powder surface by diffuse reflection infrared Fourier transform (DRIFT) spectroscopy [2]. Absorption peaks of OH related vibrational modes lie in the range 3400-3700 cm$^{-1}$. Vibrational modes of $H_3O^+$ and $H_5O_2^+$ attached to $SnO_2$ are in

the range 2990-2850, 2250-2200, 1705-1660 and 1000-900 cm$^{-1}$. Adsorption of $CO$ gives a absorption peak at 2200 cm$^{-1}$[2]. Normal vibrational modes in $H_2O$ are $\nu_1$=3657cm$^{-1}$, $\nu_2$= 1595cm$^{-1}$ and $\nu_3$=3756cm$^{-1}$ and in $CO_2$, these are $\nu_1$=1340cm$^{-1}$, $\nu_2$= 667cm$^{-1}$ and $\nu_3$=2350cm$^{-1}$ as given, in a book by Nakamoto [3].

Curves 1, 2 and 3 in fig.1(a) show PL spectra of F doped $SnO_2$ thin film deposited on glass, excited with monochromatic light from the xenon source of wavelengths 280, 330 and 380nm respectively. The inset in fig.1(a), shows normal transmission spectrum of $SnO_2$/glass thin film. Encircled peaks in fig.1(a) are due to noise of the experimental setup. Initial part of the curve1, fig.1(a), from 295 to 330nm, is due to Rayleigh scattering. The broad peak from 340 to 410 nm is due to sum of the band gap PL and PL from oxygen vacancies. The full width at half maximum (FWHM) of the band gap PL is 45nm. The maximum of the band gap PL is in the range 365-380nm. The optical band gap of SnO$_2$ at T=300K is ~345nm (3.6eV) and oxygen vacancies give PL near 390nm [4,5]. Intensities due to Rayleigh scattering for curves 2 and 3, in the range 350-450nm, are more as the intensities of the excitation wavelengths increase for longer wavelengths. Intensities due to Rayleigh scattering and intensity due to band gap PL for curves 2 and 3 are mixed. There are some sharper peaks superimposed on the PL spectrum. Intensities of superimposed peaks are maximum for the excitation wavelength 380nm. Fig.1(b) shows curves 1, 2 and 3 of fig.1(a), as a function of shift energy with respect to energies of the incident light. It is known that the intensity of RR peaks is maximum for the excitation energy equal to the sum of the band gap energy and the energy of the RR peak. In this condition, loss of energy to the lattice is minimum [6]. Let us concentrate on the curve 3 of fig.1(b), as the excitation $\lambda$ = 380nm is near the maximum of PL peak of $SnO_2$ thin film. The origins of RR peaks, observed in the present study, are assigned based on published DRIFT and FTIR results [2,3]. Possible origins of peaks near $P_1$, $P_2$, $P_3$ and $P_4$ are discussed here. The peak $P_1$ (at $2270 \pm 10$ cm$^{-1}$) of the curve 3 in fig.1(b) is probably due to the resonance Raman (RR) stretching fundamental mode of $CO$ and the peak $P_3$ (at $6970 \pm 20$ cm$^{-1}$) is due to third harmonic of stretching mode of $CO$. The peak $P_1$ (at 2270cm$^{-1}$) of the curve 3 in fig.1(b) matches with the published value of $CO$ stretching mode within experimental error and considering the dispersion of the RR peak [2]. Photoelectron spectroscopy confirmed the presence of C as impurity at the surface as well as in the bulk of $SnO_2$ thin films. Hence, it is not possible to distinguish that present results related with RR peaks are from the surface or bulk of $SnO_2$ thin film. One may also assign the origin of the peak $P_1$ to vibrational modes of hydrates ($H_3O^+$ and $H_5O_2^+$) attached to $SnO_2$. The source of peaks near $P_2$ at (3440 and $3730 \pm 20$ cm$^{-1}$) in the curve 3 in fig. 1(b) is from OH stretching modes attached differently to lattice [2]. The peak $P_4$ (at $10252 \pm 20$ cm$^{-1}$) is the third harmonic of the $OH$ stretching mode at 3440cm$^{-1}$. The RR peak positions of $CO$ and $OH$ stretching modes are at different positions for curves 1, 2 and 3 in fig.1(b) due to dispersion. Similar results are observed for $SnO_2$ thin films deposited on $Si$ substrate by the laser ablation deposition.

PL measurements were repeated on $Mo_{1-x}Fe_xO_2$ thin films, bulk fused quartz and ordinary glass slide. Curves 1 and 2 in fig. 2(a) show PL intensity as a function of shift energy for fused quartz and ordinary glass respectively for incident wavelength of 360nm. The sharp peak at $2445 \pm 10$ cm$^{-1}$ is present in both curves 1 and 2 in fig.2 (a). However, the relative intensity of this peak in the curve 2 is much smaller. The broad peak in the range 3000cm$^{-1}$ to 5500 cm$^{-1}$ in the curve 1 is absent in the curve 2 of fused quartz. This broad peak in the curve 1 of ordinary glass is due impurities of OH and other complexes related with water [3]. Curves 1 and 2 in fig. 2(b) show PL intensity as a function of shift energy for $Mo_{1-x}Fe_xO_2$ thin film for incident wavelengths 375nm and 350nm respectively. It is interesting to observe sharp peaks superimposed on PL spectra of these samples. Further, measurements were repeated for all samples by varying the incident wavelength from 260nm to 450nm to study the dispersion of RR peaks. Dispersions of the fundamental RR peak and its third harmonic are shown by curves 1 and 2 respectively in figure 3. Surprisingly, the dispersions for fundamental RR peak and its third harmonic are same for three different materials irrespective of their crystalline structures.

The slopes of the curves 1, 2 and 3 in fig.3 are 600cm$^{-1}$/eV, 800 cm$^{-1}$/eV and 2000cm$^{-1}$/eV respectively. These dispersions are much larger than reported in any other materials. The maximum reported dispersion is ~ 50 cm$^{-1}$/eV of the D peak in amorphous carbon [7,8]. The most important observation is the large dispersion of the RR peak related with *CO* and *OH* stretching modes and their higher odd harmonics. It is clear that the crystalline structure of the individual sample is not playing the major role in these observations. Impurities such as OH or CO adsorbed or trapped in defect regions at the surface and in the bulk are common to all these samples and hence, responsible for these observations. This large dispersion of RR peaks may be a common property of all oxides having defects of oxygen deficiency and OH and CO as impurities. Processes participating in RR scattering are (1) excitation of electrons from the valence bands to the conduction band (or impurity level within the band gap in the case of fused quartz and ordinary $SiO_2$ glass) and hence, creation of excess electrons and holes, (2) scattering of excess electrons and holes by vibrations of defects such as OH and CO and lattice phonons and finally (3) recombination of electrons and holes. Energy conservation requires that

$$E_i = E_s + l\hbar\omega_{lattice} + m\hbar\omega(CO, OH..) \quad \ldots\ldots\ldots \quad (1)$$

Here, $E_i$ and $E_s$ are energies of incident and scattered photons respectively, l and m are integers. $\hbar\omega_{lattice}$ is the energy absorbed by the lattice. $\hbar\omega(CO,OH..)$ are energies of vibrational modes of $CO$, $OH$ and their higher harmonics. It is well known in the case of resonance Raman that when $E_i$ is slightly more than the energy of allowed absorption between energy levels, the loss of energy to the lattice ($\hbar\omega_{lattice}$) is minimum [6]. The difference between $E_i$ and $E_s$ is almost equal to $m\hbar\omega(CO,OH..)$ when loss to lattice is negligible. Now, let us look at the dispersion of the fundamental and third harmonic of $CO$ stretching mode in fig. 2. The RR peak position of the fundamental $CO$ stretching mode is varying from 1960cm$^{-1}$ (for incident wavelength 450nm) to 3360cm$^{-1}$ (for incident wavelength 260nm). If the actual energy of the $CO$ fundamental stretching mode is ~2200cm$^{-1}$, then for RR peak positions more than 2200cm$^{-1}$, the difference is absorbed by the lattice and for RR peak positions less than 2200cm$^{-1}$ the difference is taken from the lattice. The reason for observing fundamental and its third harmonic and the absence of second and fourth, even harmonics can be understood with the help of vibrational energy levels of an asymmetric and anharmonic oscillator model for vibrations of CO and OH [3]. Transitions are allowed between symmetric and asymmetric wave-functions and vice-versa and not allowed between symmetric to symmetric or asymmetric to asymmetric wave-functions. Transitions are allowed for $\Delta v = \pm 1, \pm 3,\ldots$, hence, even harmonics are absent.

## Conclusions

Superimposed sharp peaks on the PL spectra of $SnO_2$, $Mo_{1-x}Fe_xO_2$ thin films, bulk fused quartz and ordinary $SiO_2$ glass are due to resonance Raman of vibrational modes of $CO$, $OH$ and their higher odd harmonics. The dispersions of $CO$ stretching RR modes are ~ 600 cm$^{-1}$/eV. The dispersion of the third harmonic of CO stretching mode is ~ 2000 cm$^{-1}$/eV. The dispersion of RR peaks is same for all samples irrespective of their crystalline structures. A detailed theoretical modelling is needed to explain the large dispersion of resonance Raman peaks related with fundamental stretching and higher odd harmonics of $CO$ in oxide materials with these as impurities.

**Figure Captions**

**Figure1**: (a) PL spectra of commercial F doped $SnO_2$ thin film deposited on glass at room temperature 300K. (b) PL spectra are plotted as a function of the shift energy with respect to the excitation energies. Excitation wavelengths for curves 1,2 and 3 are 280nm, 330nm and 380nm respectively. Peaks inside the circle are due to noise from the setup. The inset in (a) shows the normal transmission of the film. Resonance Raman peaks $P_1$ and $P_3$ are fundamental and third harmonic respectively of stretching mode of CO. Resonance Raman peaks $P_2$ and $P_4$ are fundamental and third harmonic respectively of stretching mode of OH.

**Figure 2**: PL spectra as a function of shift energy, curve 1 and 2 in (a) show PL spectra of fused quartz and ordinary $SiO_2$ glass for excitation wavelength 360nm. Curves 1 and 2 in (b) show PL spectra for $Mo_{1-x}Fe_xO_2$ thin film for excitation wavelengths 375nm and 350 nm respectively.

**Figure 3**: The dispersion of resonance Raman peaks related with fundamental (curve1) stretching mode of CO and its third harmonic (curve3). Symbols for three different samples $SnO_2$ thin film, $Mo_{1-x}Fe_xO_2$ thin film and ordinary $SiO_2$ glass are squares, triangles and stars respectively. The dispersion of OH fundamental stretching mode is shown by the curve 2.

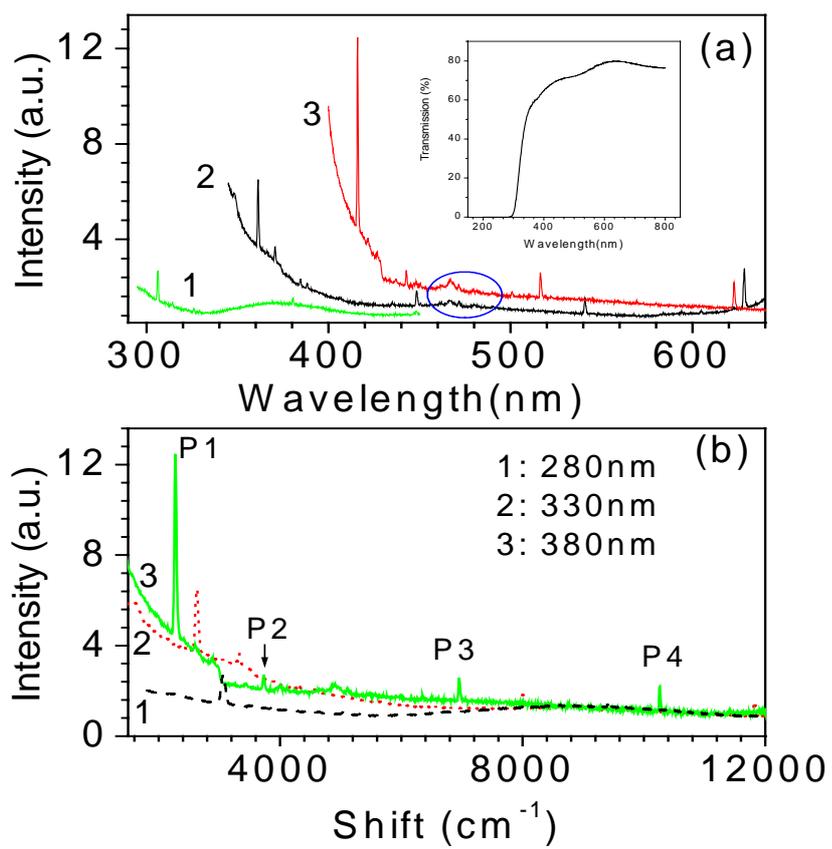

figure1

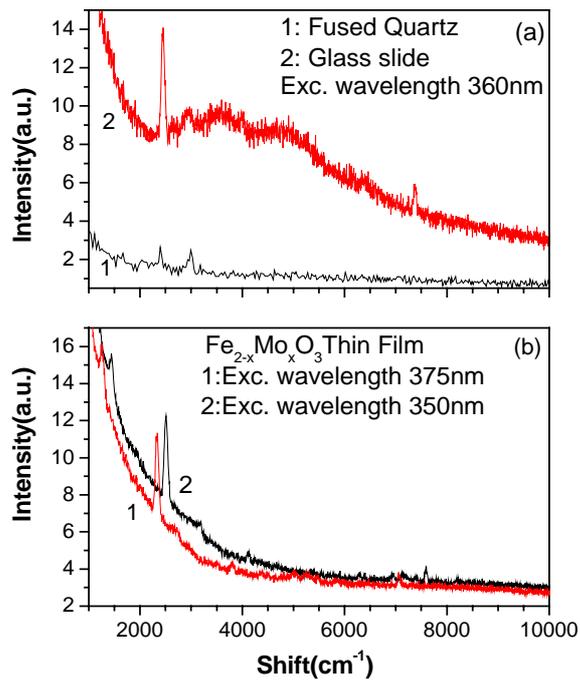

Figure 2

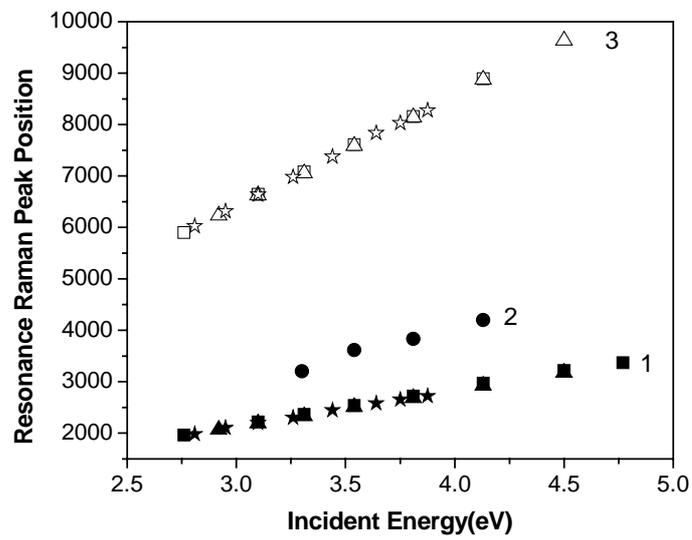

Figure 3